\documentclass[leqno,a4paper,12pt]{article}
\usepackage[centertags]{amsmath}
\usepackage{amsfonts}
\usepackage{amssymb}
\usepackage{multirow}
\usepackage{amsthm}
\usepackage{color}
\usepackage{nicefrac}
\usepackage{pifont}
\usepackage[ansinew]{inputenc}
\usepackage{graphicx}

\begin{document}

\title{An identifiability problem in a state model for partly undetected chronic diseases}
\author{Ralph Brinks}
\maketitle

\begin{abstract}
Recently, we proposed an state model (compartment model) to describe
the progression of a chronic disease with an pre-clinical (``undiagnosed'') 
state before clinical diagnosis. It is an open question, if a sequence of
cross-sectional studies with mortality follow-up is sufficient to estimate 
the true incidence rate of the disease, i.e. the incidence of the undiagnosed
and diagnosed disease. In this note, we construct a
counterexample and show that this cannot be achieved in general.
\end{abstract}

\section{Introduction}
\subsection{Compartment model}

\begin{figure}[ht]
  \centering
  \includegraphics[keepaspectratio,width=0.9\textwidth]{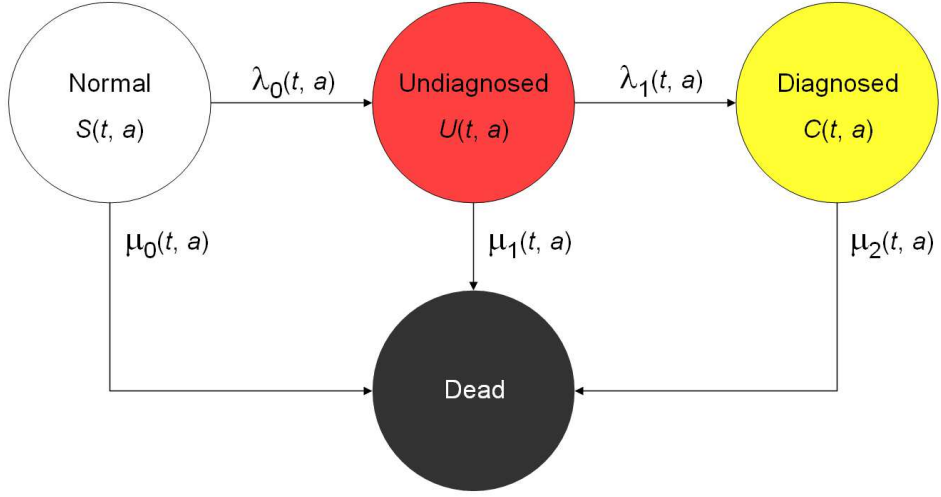}
\caption{Chronic disease model with four states and the
corresponding transition rates. People in the state \emph{Normal}
are healthy with respect to the disease under consideration. After
onset of the disease, they change to state \emph{Undiagnosed} and 
maybe later to the state \emph{Diagnosed}. The absorbing state \emph{Dead} can be
reached from all other states. The numbers of persons in the states and
the transition rates depend on calendar time $t$ and age $a.$}
\label{fig:CompModel}
\end{figure}

Recently, we introduced a compartment model with a pre-clinical stage
preceding the clinical stage \cite{Bri15b}. The model involves
calendar time $t$, and the different ages $a$ of the 
subjects in the population. The transition rates between the
states are denoted as in Figure \ref{fig:CompModel}.

Using the definition $N(t, a) = S(t, a) + U(t, a) + C(t, a)$ and setting
\begin{align*}
p_0(t, a) &= \frac{S(t, a)}{N(t, a)} \\
p_1(t, a) &= \frac{U(t, a)}{N(t, a)} \\
p_2(t, a) &= \frac{C(t, a)}{N(t, a)},
\end{align*}

the compartment model in Figure \ref{fig:CompModel} is governed
by a system of partial differential equations (PDEs):

\begin{align}
(\partial_t + \partial_a) p_1 &= - \bigl ( \lambda_0 + \lambda_1 + \mu_1 - \mu \bigr ) \, p_1 - \lambda_0 \, p_2 + \lambda_0 \label{e:pde1}\\
(\partial_t + \partial_a) p_2 &= \lambda_1 \, p_1 - \bigl ( \mu_2 - \mu \bigr ) \, p_2 \label{e:pde2}.
\end{align}

For brevity we have written $\partial_x = \frac{\partial}{\partial x}, ~x\in \{t, a\}.$
In Eq. \eqref{e:pde1} -- \eqref{e:pde2} the general mortality $\mu$ is given
by 
\begin{equation}\label{e:genMort}
\mu = p_0 \mu_0 + p_1 \mu_1 + p_2 \mu_2.
\end{equation}

Together with the initial conditions 
$p_1(t , 0) = p_2(t , 0) = 0$ for all $t,$ the
system \eqref{e:pde1} -- \eqref{e:pde2} completely describes the temporal 
dynamics of the disease in the considered population. The quantity $p_0$
can be obtained by 
\begin{equation}\label{e:p0}
p_0 = 1 - p_1 - p_2.
\end{equation}

\subsection{Direct and inverse problem}

Assumed that the functions $\lambda_0, \lambda_1, \mu_1, \mu_2, \mu$ on the right-hand
sides of system \eqref{e:pde1} -- \eqref{e:pde2} are suffiently smooth, then
the associated initial value problem $p_1(t , 0) = p_2(t , 0) = 0$ for all $t$
has a unique solution. This means that together with the initial condition,
there is a function 
\begin{equation}\label{e:opera}
\Phi: \Theta = \bigl ( \lambda_0, \lambda_1, \mu_1, \mu_2, \mu \bigr ) \mapsto P = (p_1, p_2).
\end{equation}

Given the initial conditions, the operator $\Phi$
maps the transition rates $\Theta$ onto the uniquely associated 
prevalence functions $\Phi(\Theta) = P = (p_1, p_2).$ This problem is called
the \emph{direct problem} or \emph{forward problem} \cite{Ast11}.

Similar to the simpler compartment model in \cite{Bri15}, the
question arises if and under which circumstances the opposite way is
possible. Does a series prevalence studies $P$ allow to estimate the 
transition rates $\Theta$? Mathematically, this problem is expressed as
inversion of the function $\Phi$. Given $P,$ the question is if
there is a unique $\Theta$ such that $\Phi(\Theta) = P?$ 
The problem of estimating the rates from prevalence data, is called 
an \emph{inverse problem} \cite{Ast11}. It is not guaranteed that the
inverse problem has a solution. Examination
of conditions such that the inverse problem has a solution is
called the analysis of \emph{identifiability} \cite{Eis86}.

Under certain circumstances, the operator $\Phi$ is indeed invertible.
Assumed that the mortality rates $\mu_1, \mu_2,$ and $\mu$ are known,
then for given $P = (p_1, p_2)$ the system \eqref{e:pde1} -- \eqref{e:pde2} 
can be solved for $\lambda_0$ and $\lambda_1.$ Thus, in these cases
$\Phi$ is invertible.

In the next section, we will show that is not always the case.


\section{Identifiability problem}

We consider two prevalence studies at
calendar times $t_1 < t_2$ with \emph{mortality follow-up.} This means, on the one hand 
we have estimates for the age courses of the prevalences $p_1$ and $p_2$
at $t_1$ and $t_2.$ On the other hand, we have additional information
if and when any participant at $t_1$ has died before $t_2.$

Let us assume that for any participant who deceased between $t_1$ and $t_2$,
we do not have information about what state the person was in at 
the time of death. For example, a person
who was in the \emph{Normal} state at $t_1$ and died before $t_2$
could have deceased when he was still in the \emph{Normal} state, in
the \emph{Undiagnosed} state or in the \emph{Diagnosed} state. 
An exception is someone dying between $t_1$ and $t_2$, 
who was in the \emph{Diagnosed} state.
As the \emph{Diagnosed} state can only be left via the transition to 
\emph{Dead} state, the information from the mortality follow-up helps
to estimate $\mu_2.$ Thus,
the mortality follow-up contributes to estimate
the general mortality $\mu$ or occasionally the mortality $\mu_2$, 
but not to estimate $\mu_0$ or $\mu_1.$

\bigskip

The question arises: Given $p_k(t_j, \cdot), ~j, k=1, 2, ~\mu(t^\star, \cdot)$
and $\mu_2(t^\star, \cdot)$ for some $t^\star$ with $t_1 < t^\star < t_2,$
are we able to estimate the rates $\lambda_0, \lambda_1, \mu_0,$ and $\mu_1$ at $t^\star?$
In the following we will show that this is not the case. This is done by
constructing a counterexample with given $p_1, p_2, \mu, \mu_2$ but different
$\lambda_0, \lambda_1, \mu_0,$ and $\mu_1.$

\bigskip

Consider the system \eqref{e:pde1} -- \eqref{e:pde2} being in equilibrium such that 
$\partial_t \, p_k(t^\star, a) = \partial_a \, p_k(t^\star, a) = 0, ~k=1,2,$ for all $a.$
Furthermore, let $p_0 = 0.5, p_1 = 0.3$ and $p_2 = 0.2, \mu = 0.6, \mu_2 = 0.8.$
Obviously, it holds $p_0 + p_1 + p_2 = 1.$ From $\partial_x p_2 = 0, ~x\in \{t, a\}$ 
it follows that $\lambda_1 = (\mu_2 - \mu) \tfrac{p_2}{p_1} = \tfrac{4}{30}.$
If we choose $\mu_1^{(1)} = 0.5$ and 
$\mu_1^{(2)} = 0.6,$ then from $\mu = p_0 \mu_0 + p_1 \mu_1 + p_2 \mu_2$ it follows
that $\mu_0^{(1)} = 0.58$ and $\mu_0^{(2)} = 0.52.$ In addition, $\partial_x p_1 = 0, 
~x\in \{t, a\}$
implies $\lambda_0 = (\lambda_1 + \mu_1 - \mu) \tfrac{p_1}{p_0}.$ Thus, it holds
$\lambda_0^{(1)} = 0.02$ and $\lambda_0^{(2)} = 0.08.$ The results are summarized
in Table \ref{tab:res}.

\begin{table}[ht]
     \centering
     \begin{tabular}{c|cc}
       Variable  & Value 1 & Value 2        \\ \hline
       $p_0$              & \multicolumn{2}{|c}{0.5} \\
       $p_1$              & \multicolumn{2}{|c}{0.3} \\
       $p_2$              & \multicolumn{2}{|c}{0.2} \\
       $\mu$              & \multicolumn{2}{|c}{0.6} \\
       $\mu_2$            & \multicolumn{2}{|c}{0.8} \\ \hline
       $\lambda_1$        & \multicolumn{2}{|c}{$\nicefrac{4}{30}$} \\
       $\mu_1$            & 0.5  & 0.6               \\
       $\mu_0$            & 0.58 & 0.52              \\
       $\lambda_0$        & 0.02 & 0.08              \\
     \end{tabular}

     \caption{Example for non-identifiability of the system \eqref{e:pde1} -- \eqref{e:p0}.
     In an equilibrium state ($\partial_x p_k = 0, ~k=1,2, ~x\in \{t, a\}$), 
     measured values in the upper
     half of the table are consistent with the values in the lower half.}
     \label{tab:res}
\end{table}

Hence, from given $p_1, p_2, \mu, \mu_2,$ in equilibrium, we were able to 
construct \emph{different}
$\lambda_0, \lambda_1, \mu_0,$ and $\mu_1,$ which are consistent with the system
\eqref{e:pde1} -- \eqref{e:p0}. This implies that two cross-sections at $t_1$ and $t_2$
with mortality follow-up are not sufficient to make the system identifiable.


\section{Conclusion}
In this technical note it was shown by a counterexample that two cross-sectional
studies with mortality follow-up are not sufficient to make the system 
\eqref{e:pde1} -- \eqref{e:p0} identifiable. This means, from two cross-sectional
studies and measured $p_k, ~k=0,1,2,$ and known $\mu, \mu_2$ it is not possible to
estimate the incidence rates $\lambda_0$ and $\lambda_1.$

\bigskip

The counterexample was constructed by the system \eqref{e:pde1} -- \eqref{e:pde2}
being in equilibrium. This is not a loss of generalizability. 
It is sufficient to find one example of non-identifiability
to prove non-existence of a solution of the inverse problem.

\bigskip

Note that from measured $p_k, ~k=0,1,2,$ and known $\mu, \mu_2,$ the rate $\lambda_1$
is estimable. This can be seen by solving Eq. \eqref{e:pde2} for $\lambda_1.$


\bibliography{references}

\begin{thebibliography}{1}

\bibitem{Bri15b}
Brinks R, Bardenheier BH, Hoyer A, Lin J, Landwehr S, Gregg EW.
\newblock Development and demonstration of a state model for the estimation of
  incidence of partly undetected chronic diseases.
\newblock BMC Medical Research Methodology. 2015;15(1):98.

\bibitem{Ast11}
Aster RC, Borchers B, Thurber CH.
\newblock Parameter estimation and inverse problems.
\newblock Academic Press; 2011.

\bibitem{Bri15}
Brinks R, Landwehr S.
\newblock A new relation between prevalence and incidence of a chronic disease.
\newblock Mathematical Medicine and Biology. 2015;.

\bibitem{Eis86}
Eisenfeld J.
\newblock A simple solution to the compartmental structural-indentifiability
  problem.
\newblock Mathematical Biosciences. 1986;79(2):209--220.

\end{thebibliography}

\emph{Contact:} \\
Ralph Brinks \\
German Diabetes Center \\
Auf'm Hennekamp 65 \\
D- 40225 Duesseldorf\\
\verb"ralph.brinks@ddz.uni-duesseldorf.de"
\end{document}